\documentclass[showpacs,preprintnumbers,amsmath,amssymb,twocolumn]{revtex4}

\usepackage{graphicx}

\begin{document}

\title{Dynamics of Spin Relaxation near the Edge of Two-Dimensional Electron Gas}

\author{Yuriy V. Pershin}
\affiliation{Center for Quantum Device Technology,\\ Department of
Physics, Clarkson University, Potsdam, NY 13699-5721, USA}

\begin{abstract}
We report calculations of spin relaxation dynamics of
two-dimensional electron gas with spin-orbit interaction at the
edge region. It is found that the relaxation of spin polarization
near the edge is more slow than relaxation in the bulk. That
results finally in the spin accumulation at the edge. Time
dependence of spin polarization density is calculated analytically
and numerically. The mechanism of slower spin relaxation near the
edge is related to electrons reflections from the boundary and the
lack of the translation symmetry. These reflections partially
compensate electron spin precession generated by spin-orbit
interaction, consequently making the spin polarization near the
edge long living. This effect is accompanied by spin polarization
oscillations and spin polarization transfer from the perpendicular
to in-plane component.
\end{abstract}


\maketitle

The prospects for creating a semiconductor-based spintronic device
\cite{book,review,datta,Per12,shlim0,Vagner} have generated an
emphasis on the studies of properties of electron spin
polarization in semiconductor nanostructures
\cite{Per20,Wu,Malsh0,Kisel,Per17,Malsh1,Per21,Bruno,sherman,puller,Sayka}.
Great interest has been expressed in dynamics of electron spin
polarization \cite{Per20,Wu,Malsh0,Kisel,Per17,Malsh1}. For
instance, spin relaxation dynamics has been studied in
two-dimensional electron gas (2DEG) \cite{Per20,Wu},
two-dimensional channels \cite{Kisel,Malsh0}, open Sinai billiards
(2DEG with a lattice of antidots) \cite{Per17}, and ballistic
quantum dots \cite{Malsh1}. It was found that the sample geometry
\cite{Per17} as well as specific initial conditions
\cite{Per20,Wu} could have a significant effect on electron spin
relaxation.

D'yakonov-Perel' (DP) spin relaxation mechanism \cite{DP} is the
leading spin relaxation mechanism in many important experimental
situations. In the framework of DP theory, initially homogeneous
electron spin polarization exponentially relaxes to zero (or to
some finite equilibrium value) with time. However, DP theory was
formulated for the bulk of a sample. Considering electron spin
relaxation near the edge of 2DEG, one would expect the same
relaxation scenario. This expectation, however, is not correct. We
demonstrate in this Letter that the spin relaxation dynamics near
the edge is rather unusual and can not be described by a simple
exponential law, as follows from the DP theory. We observe a
longer spin relaxation time near the edge, spin polarization
oscillations and spin polarization transfer from the perpendicular
(to 2DEG) to in-plane component.

Let us consider a two-dimensional electron gas with the Rashba
spin-orbit interaction \cite{Rashba}, which couples electron space
and spin degrees of freedom:

\begin{equation}
H_R=\alpha \hbar^{-1}\left( \sigma_x p_y -\sigma_y p_x \right),
\label{RashbaHam}
\end{equation}
where $\alpha$ is a constant, $\mathbf{\sigma}$ is the Pauli
matrix vector corresponding to the electron spin, and $\mathbf{p}$
is the momentum of the electron confined in two-dimensional
geometry. From the point of view of electron spin, the effect of
spin-orbit interaction can be regarded as an effective magnetic
field acting on electron spin. Momentum scatterings reorient the
direction of this field, thus leading to average spin relaxation
(DP relaxation). Intuitively, electron reflections from the 2DEG
edges should increase spin relaxation time, since electron spin
rotations will be partially compensated. In order to study this
phenomena, we use both analytical and numerical approaches.

\begin{figure}[b]
\centering
    \includegraphics[width = 8cm]{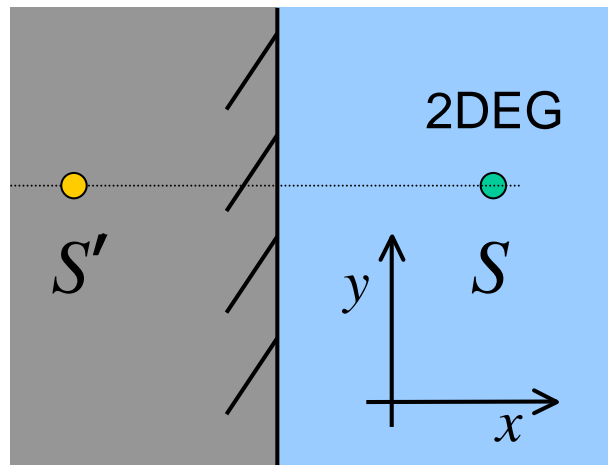}
    \caption{Schematic of the system: 2DEG occupies half-plane $x>0$. The
    influence of the boundary located at $x=0$ on the electron space motion is taken into account
    by introducing the mirror source $S'$.}
\end{figure}
Let us consider a two-dimensional electron gas occupying
half-space $x>0$ (Fig. 1) with electron spins homogeneously
polarized in $z$ direction, perpendicular to 2DEG, in the initial
moment of time $t=0$. From the experimental point of view, the
initial electron spin polarization as well as detection of spin
polarization can be realized using the optical technique
\cite{Awsch} or any other appropriate method. It is convenient to
describe the spin state of $i$-th electron via the spin
polarization vector $\mathbf{P}_i=Tr(\rho_i\mathbf{\sigma})$,
where $\rho_i$ is the single-particle density matrix \cite{blum}.
In the framework of drift-diffusion approximation, the initial
dynamics of the spin polarization density $\mathbf{P}$, which is
defined via $\mathbf{P}d^2r=\sum\limits_{i\in d^2r}\mathbf{P}_i$
(here $d^2r=dxdy$ is a space volume), can be calculated using the
following expression

\begin{equation}
\mathbf{P}(x,y,t)=\int\limits_0^\infty\int\limits_{-\infty}^\infty
G(x-x',y-y',t)\mathbf{P'}_{(x,y),(x',y')}dy'dx', \label{MainEq}
\end{equation}
where $G(x-x',y-y',t)$ is the diffusion Green function (solution
of a diffusion equation with a point source), and
$\mathbf{P'}_{(x,y),(x',y')}$ represents a contribution of the
initial spin polarization density at point $(x',y')$ to
$\mathbf{P}(x,y,t)$. The structure of Eq. (\ref{MainEq}) can be
easily understood. Electron spin polarization density in a space
volume with coordinates $(x,y)$ at a selected moment of time $t$
is given by a sum of spin polarization vectors of all electrons
located in this volume. The diffusion Green function
$G(x-x',y-y',t)$ gives probability for electrons to diffuse from
the point $(x',y')$ to $(x,y)$, while
$\mathbf{P'}_{(x,y),(x',y')}$ describes the spin polarization of
these electrons.

We note that Eq. (\ref{MainEq}) governs only the initial spin
relaxation dynamics. The main approximation behind Eq.
(\ref{MainEq}) is the assumption that different spin rotations
commute with each other, and spin precession angle $\varphi$ is
proportional to the distance between $(x',y')$ and $(x,y)$. This
assumption is justified when the spin precession angle per mean
free path is small and the time is short. Moreover, it is assumed
that evolution of electron spin degree of freedom is superimposed
on the space motion of the charge carriers. In other word, the
influence of the spin-orbit interaction on electron space motion
is neglected. If $\mathbf{a}$ is the unit vector along the
precession axis, then \cite{Per21}

\begin{equation}
\mathbf{P'}_{(x,y),(x',y')}=\mathbf{P}+\mathbf{P}_\bot\left( \cos
\varphi -1 \right)+ \mathbf{a}\times\mathbf{P}\sin \varphi,
\label{P}
\end{equation}
where $\mathbf{P}_\bot=\mathbf{P}-\mathbf{a}
(\mathbf{a}\mathbf{P})$ is the component of the spin polarization
perpendicular to the precession axis, $\varphi=\eta r$, $\eta$ is
the spin precession angle per unit length,
$\mathbf{r}=(x-x',y-y')$, $r=|\mathbf{r}|$,
$\mathbf{a}=\hat{z}\times\mathbf{r}/r$, and $\hat{z}$ is the unit
vector in $z$ direction. We emphasize that the Rashba spin-orbit
interaction is the origin of spin polarization rotations described
by Eq. (\ref{P}). The coupling constant $\alpha$ of the Rashba
interaction enters into our model through the parameter $\eta$
(for details see Ref. \cite{Kisel}).

When 2DEG occupies only a half-space, the Green function
$G(x-x',y-y',t)$ that appears in Eq. (\ref{MainEq}) can be
obtained using mirror-image approach. Within this approach, the
Green function $G(x-x',y-y',t)$ is written as a superposition of
two point-source solutions

\begin{equation}
G=G_0(x-x',y-y',t)+G_0(x-x',y+y',t), \label{GF}
\end{equation}
where $G_0(\mathbf{r},t)=1/(4 \pi D t)e^{-\mathbf{r}^2/(4Dt)}$ is
the full space Green function, $D=L_p^2/(2\tau_p)$ is the
diffusion coefficient, $L_p$ is the mean free path, and $\tau_p$
is the momentum relaxation time. While the first term in the right
hand side in Eq. (\ref{GF}) describes the real source, the second
term is a so-called mirror source (Fig. 1), that is introduced to
describe the reflections of electrons from the boundary.

\begin{figure}[t]
\centering
    \includegraphics[width = 8cm]{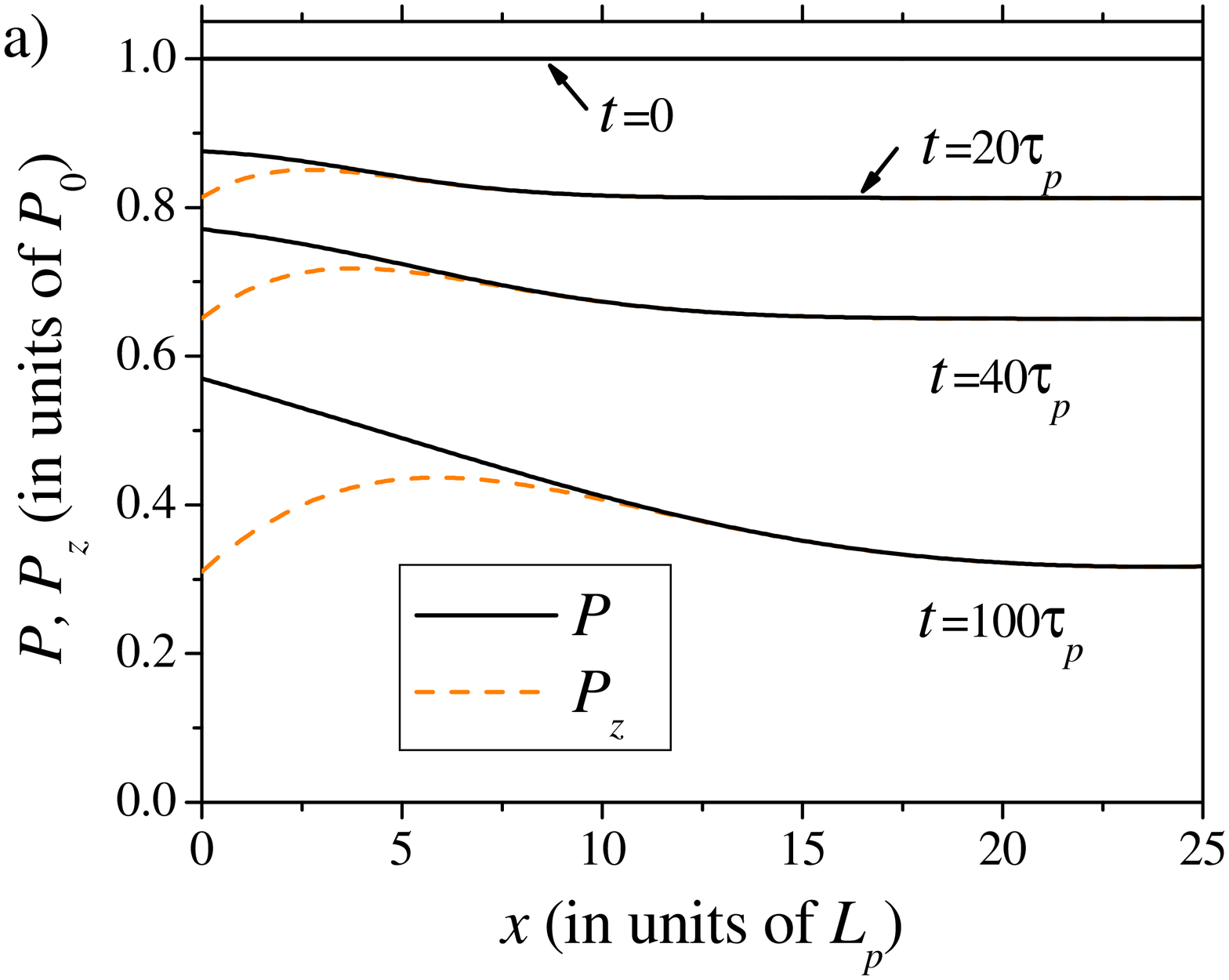}
    \includegraphics[width = 8cm]{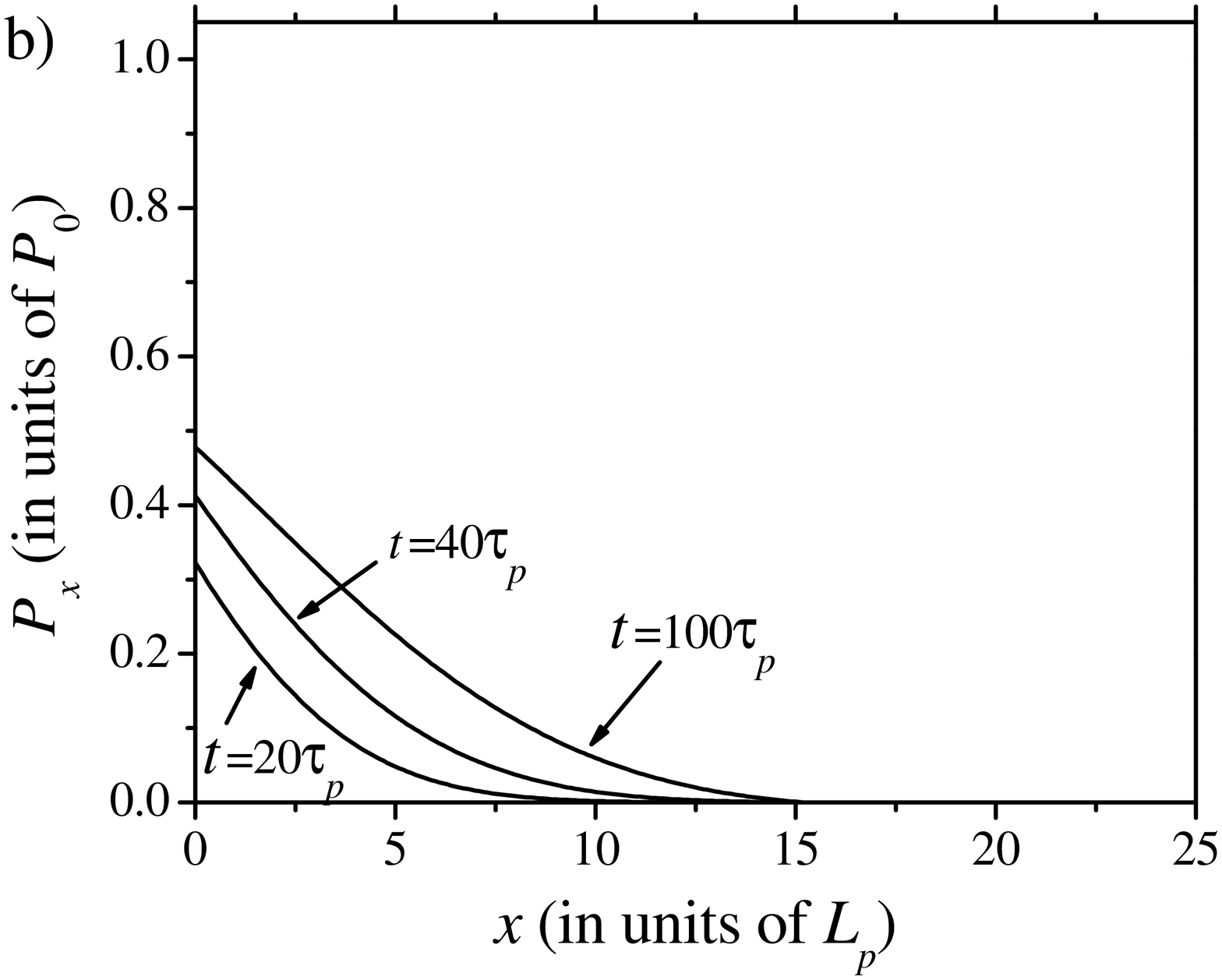}
    \caption{ (Color online) Initial evolution of spin polarization near the edge calculated within
    the drift-diffusion approximation. Relaxation of electron spins occurs
    slowly near the edge. All graphs presented in this Letter were obtained using the same parameter $\eta L_p=0.1$.}
\end{figure}

Substituting Eq. (\ref{P}) into Eq. (\ref{MainEq}) and taking the
initial spin polarization in the form $\mathbf{P}(t=0)=P_0
\hat{z}$, we can obtain the following expressions for the
components of the spin polarization density:

\begin{equation}
 P_x= P_0\int\limits_0^\infty\int\limits_{-\infty}^\infty
G(x-x',y-y',t)\frac{x'-x}{r}\sin(\eta r)dy'dx'
\label{Px}\end{equation}
\begin{equation} P_y=0 \label{Py} \end{equation} \begin{equation}
P_z=P_0\int\limits_0^\infty\int\limits_{-\infty}^\infty
G(x-x',y-y',t)\cos(\eta r)dy'dx' \label{Pz}.
\end{equation}
It is obvious that $P_y=0$ from symmetry considerations. We can
not perform the integration in Eqs. (\ref{Px}) and (\ref{Pz}) in
the general form. However, these integrals can be evaluated
explicitly for short times. In this case, since the exponents of
the Green function effectively cut off contributions of the terms
with large $r$, the geometric functions in Eqs. (\ref{Px}),
(\ref{Pz}) can be approximated as $\sin(\eta r)\approx \eta r $
and $\cos(\eta r)\approx 1-(\eta r)^2/2$. Then, after the
integration, we get

\begin{equation}
\frac{P_x}{P_0}=\eta x \left( 1-Erf\left[
\frac{x}{\sqrt{4Dt}}\right]\right) - \eta
\sqrt{\frac{4Dt}{\pi}}e^{-\frac{x^2}{4Dt}} \label{Px1}
\end{equation}
and
\begin{eqnarray}
\frac{P_z}{P_0}=1-\eta^2 x^2 \left( 1-Erf\left[
\frac{x}{\sqrt{4Dt}}\right]\right) + \nonumber \\ 2Dt
\eta^2\left(-1+ \frac{xe^{-\frac{x^2}{4Dt}}}{\sqrt{\pi
Dt}}\right), \label{Pz1}
\end{eqnarray}
where, $Erf\left[z\right]$ is the error function, given by
$Erf\left[z\right]=\left( 2/\sqrt{\pi}
\right)\int\limits_0^{z}e^{-t^2}dt$.  It follows from Eqs.
(\ref{Px1}) and (\ref{Pz1}) that the area of nonhomogeneous spin
polarization density near the edge increases in time and is
characterized by the length $l \sim \sqrt{4Dt}$. The behavior of
spin relaxation in the bulk region is governed by the standard DP
theory: $P_x=0$ and the relaxation of $P_z$ is described by the DP
relaxation time $\tau_r=\tau_p/ (L_p \eta)^2$. Let us estimate the
limits of applicability of Eqs. (\ref{Px1}), (\ref{Pz1}). The
characteristic length scale of the Green functions is
$\sqrt{4Dt}$, while the characteristic length of the geometric
functions is $\eta^{-1}$. Setting these two lengths equal we find
that  Eqs. (\ref{Px1}), (\ref{Pz1}) are valid while $t \lesssim
1/(4D\eta^2)$.

Figure 2 shows the spin polarization density
$P=\sqrt{P_x^2+P_z^2}$ and its components $P_x$ and $P_z$ ($P_y=0$
accordingly to Eq. (\ref{Py})) as a function of $x$ at several
moments of time. Spin polarization density represented in Fig. 2
was obtained using a numerical integration in Eqs. (\ref{Px}) and
(\ref{Pz}). It is clearly seen the absolute value of the spin
polarization density $P$ is higher in the edge region. It is
interesting to note the the higher spin polarization density near
the edge is largely due to the $x$ component of spin polarization
density (Fig. 2b). From Eq. (\ref{Px}) it follows that initial
growth of $P_x$ at $x=0$ is proportional to $\sqrt{t}$. We also
observe formation of a local maximum of $P_z$ in the vicinity of
the edge. The peak of $P_z$ drifts from $x=0$ with time.

\begin{figure}[t]
\centering
    \includegraphics[width = 8cm]{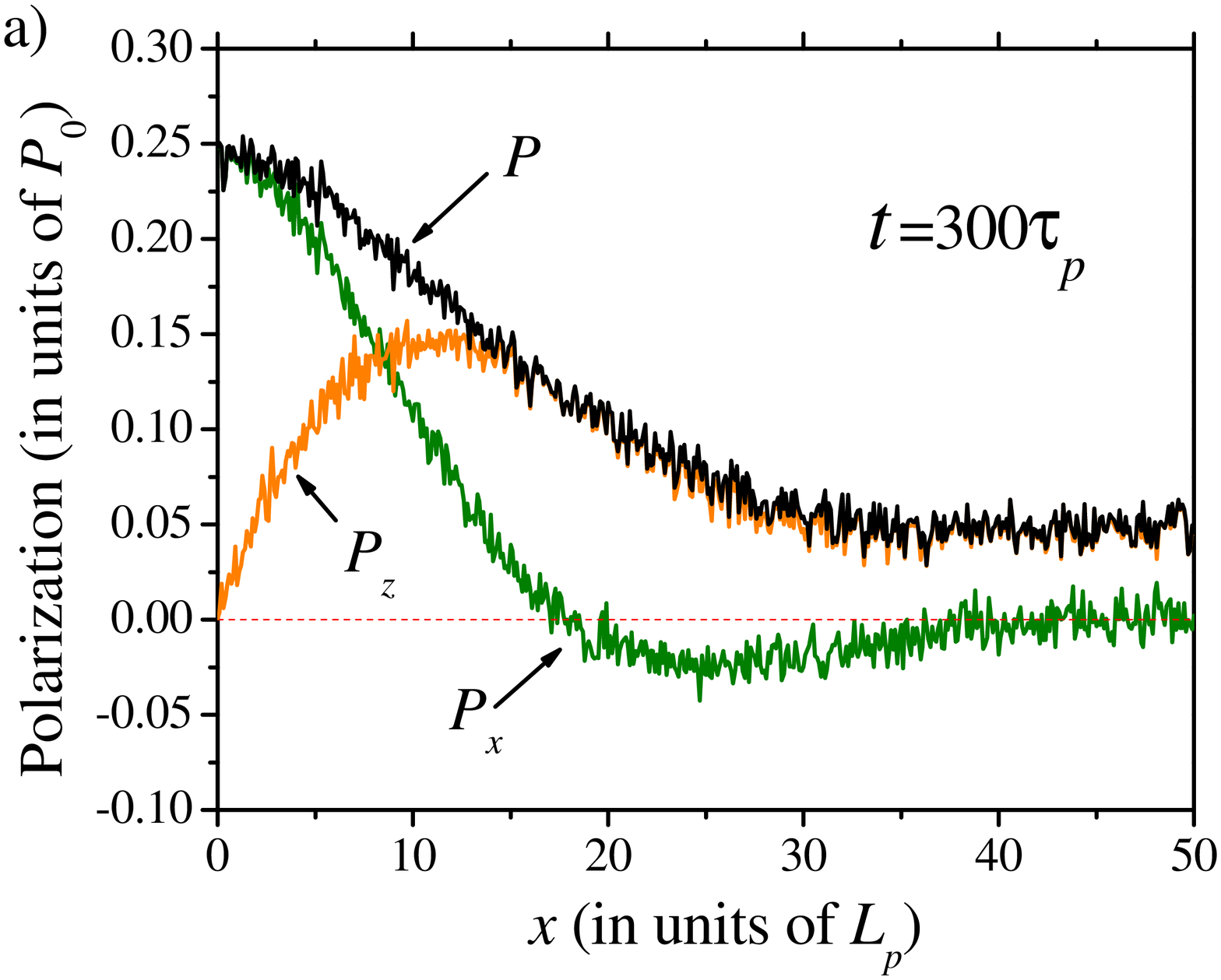}
    \includegraphics[width = 8cm]{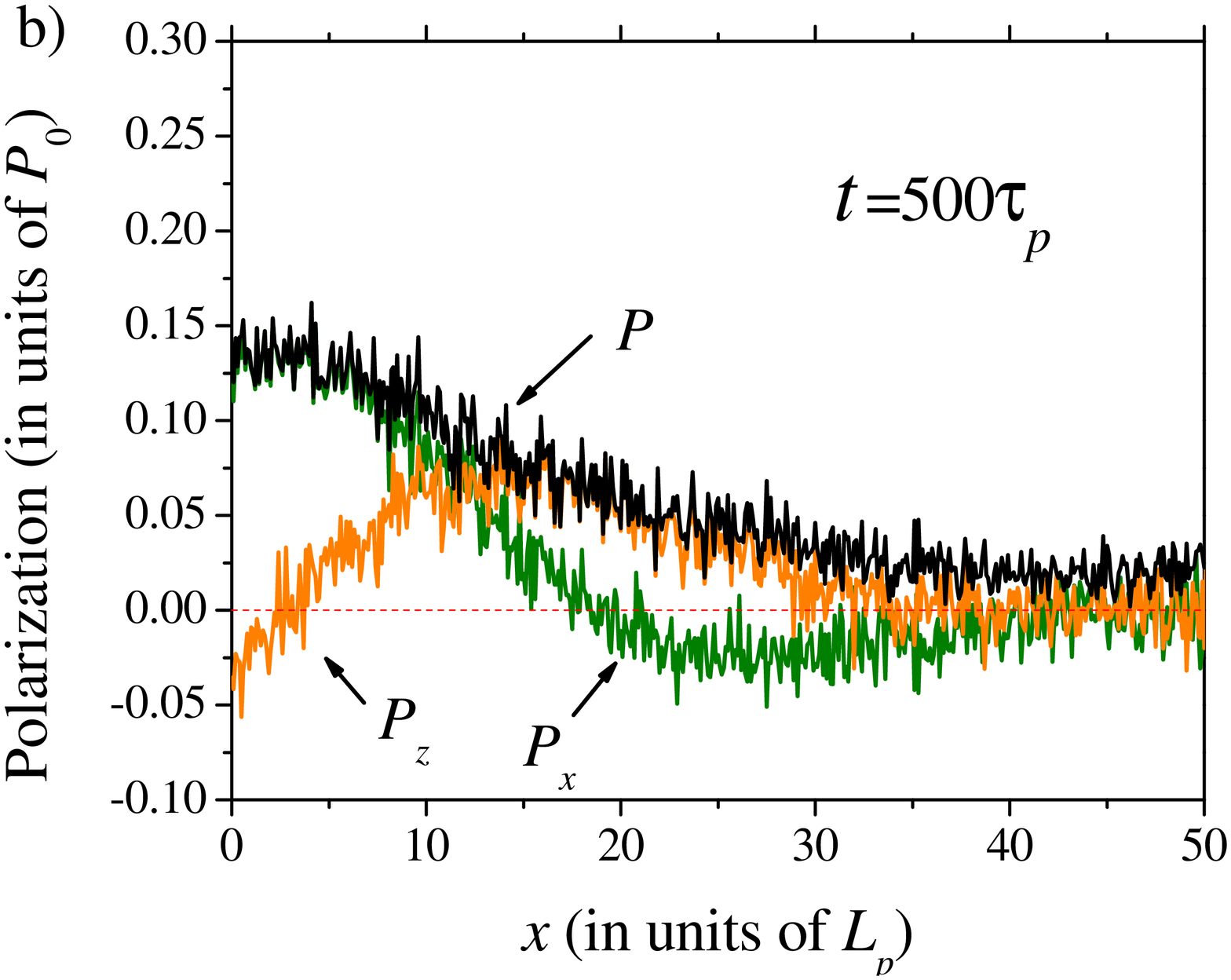}
    \caption{ (Color online) Spin polarization near the edge at longer times.
    Suppression of spin relaxation near the edge, as well as spin oscillations
    in time and space are observed. The dotted red line is for the eye.} \label{MCres}
\end{figure}

In order to check the results of our drift-diffusion approximation
as well as to find long time behavior of spin polarization
density, we study the dynamics of spin relaxation near the edge
using a Monte Carlo simulation scheme that was proposed in Ref.
\cite{Kisel} and subsequently used in Refs. \cite{Per17,Per20}.
Within the Monte Carlo simulation algorithm, the space motion of
2DEG electrons is considered to be along classical (linear)
trajectories interrupted by the bulk scattering events. Our
modeling involves spin-independent bulk scattering processes,
which could be caused, e.g., by impurities. For the sake of
simplicity, the scattering due to such events is assumed to be
elastic and isotropic, i.e., the magnitude of the electron
velocity is conserved in the scattering, while the final direction
of the velocity vector is randomly selected. The time scale of the
bulk scattering events can then be fully characterized by a single
rate parameter \cite{Kisel}, the momentum relaxation time,
$\tau_p$. It is connected to the mean free path by $L_p=v \tau_p$.
Here $v$ is the mean electron velocity. We use reflecting boundary
conditions from the edge: the longitudinal component of the
electron velocity is preserved and the sign of the normal
component is changed in collisions. The typical number of
electrons used in our simulations is $2\cdot 10^7$.

We have found that the results of Monte Carlo simulations at short
times are in perfect agreement with predictions of drift-diffusion
model. Fig. \ref{MCres} shows spin polarization density obtained
via Monte Carlo simulations at longer times. At these times the
space distribution of spin polarization is nontrivial and the
slowing down of the spin relaxation near the edge becomes more
pronounced. Fig. \ref{MCres} indicates the change of the sign of
$P_z$ at $x=0$. The change of the sign of $P_x$ occurs some
distance away from the edge. Relaxation of spin polarization
density in the bulk is in agreement with DP theory.

In conclusion, we have studied the spin relaxation dynamics in a
2DEG near the edge. The obtained result indicates that the spin
relaxation in the edge region deviates from the D'yakonov-Perel'
theory. Specifically, spin relaxation occurs slowly in the edge
region and is accompanied by spin oscillations in time and space
and spin polarization transfer from perpendicular to in-plane
component. We have derived an analytical formula describing
initial spin polarization dynamics. Long time behaviour of spin
polarization density was studied numerically. Qualitatively, the
observed effects are related to the lack of translation symmetry
in the system. Reflections of electron from the boundary partially
compensate spin precessions due to spin-orbit interaction.
Moreover, the lack of the actual electron flow from the half-plane
$x<0$ gives rise to non-zero in-plane component of spin
polarization density near the edge, since this component of spin
polarization for electrons coming from $x>0$ into the edge region
is uncompensated. In some sense the structure that forms near the
edge is similar to long-living spin coherence states previously
discussed in Ref. \cite{Per20}. We note that our results are not
directly related to the spin Hall effect \cite{SpinHall}, in which
the electron spins become polarized in an applied electric field.
Nevertheless, our results could be important in interpretation of
spin Hall experiments.

We acknowledge useful discussions with Prof. V. Privman. This
research was supported by the National Security Agency and
Advanced Research and Development Activity under Army Research
Office contract DAAD-19-02-1-0035, and by the National Science
Foundation, grant DMR-0121146.

\end{document}